\documentclass[12pt,preprint]{aastex}
\usepackage{rotating}

\slugcomment{To be submitted to the Astrophysical Journal}
\shorttitle{Evidence of the FIP effect in the coronae of late-type giants}
\shortauthors{Garc\'{\i}a-Alvarez et al.}
\begin{document}
\title{Evidence of the FIP effect in the coronae of late-type giants}
\author{David Garc\'{\i}a-Alvarez, Jeremy J. Drake, B. Ball, LiWei Lin, Vinay L. Kashyap}
\affil{$^1$Harvard-Smithsonian CfA, \\ 60 Garden Street, \\ Cambridge, MA 02138}

\begin{abstract}
$\beta$ Cet, 31 Com and $\mu$ Vel represent the main stages through
which late-type giants evolve during their lifetime (the Hertzsprung
gap (31 Com), the rapid braking zone ($\mu$ Vel) and the core helium
burning ``clump'' phase ($\beta$ Cet)). An analysis of their high 
resolution {\it Chandra} X-ray spectra reveals similar coronal 
characteristics in terms of both temperature structure and element 
abundances for the more evolved stars ($\mu$ Vel and $\beta$ Cet) with 
slight differences for the `younger' giant (31 Com). The coronal 
temperature structure of 31 Com is significantly hotter showing a clear peak while $\beta$ Cet 
and $\mu$ Vel show a plateau. $\beta$ Cet and $\mu$ Vel show evidence 
for a FIP effect in which coronae are depleted in high FIP elements
relative to their photospheres by a factor of $\sim 2$. In contrast, 
31~Com is characterized by a lack of FIP effect. In other words, neither
depletion nor enhancement relative to stellar photospheric values is
found. We conclude that the structural changes during 
the evolution of late-type giants could be responsible for the observed 
differences in coronal abundances and temperature structure. In particular, 
the size of the convection zone coupled with the rotation rate seem obvious 
choices for playing a key role in determining coronal characteristics. 

\end{abstract}


\section{Introduction}

The solar coronal abundance anomaly commonly known as the ``FIP
Effect'' (or First Ionization Potential Effect), in which low first
ionization potential (FIP) elements (e.g.; Si, Fe, Mg) appear enhanced
by average factors of 3-4, was already { known} by the time the first
observational clues to similar abundance anomalies in stellar coronae
were uncovered in the 1990's \citep[e.g.;][]{Feldman92}.  These clues
came from low resolution soft X-ray spectra ({\it GINGA, ASCA,
BeppoSAX}), together with moderate resolution extreme ultraviolet
(EUV) spectra obtained by the Extreme Ultraviolet Explorer (EUVE). The
early stellar studies found evidence for a solar-like FIP Effect in
some stars, but for the more active stars the abundances pointed
toward metal paucity rather than enhancement \citep{Drake94,White94}.

The last four years---the beginning of the {\it Chandra} and {\it
XMM-Newton} era--- have seen early hints of abundance anomalies
fleshed out into an interesting array of diverse abundance patterns in
which active stars appear to show signs not only of low FIP element
depletion, but also of high FIP element enhancements.  Studies of
solar-like active stars confirm the suspicions engendered by earlier
work \citep[e.g.;][]{Drake95} that a transition from a metal-depleted
to a metal-rich corona occurs as the activity decreases
\citep{Guedel02}; this is now better characterised as a change from an
apparently ``inverse FIP effect'' to a FIP effect. \citet{Audard03}
showed a similar transition from an inverse FIP effect to the absence
of an obvious FIP effect for RS CVn binary stars with decreasing
activity. \citet{Drake95}, \citet{Raassen02} and
\citet{Sanz-Forcada04} showed that coronal abundances can also be
similar to that of its underlying photosphere. \citet{Sanz-Forcada04}
also presented evidence of possible variation in coronal metal
abundance relative to the temperature of the emitting plasma.


For some years we have argued \citep[e.g.,][ and earlier references
therein]{Drake03a} that coronal abundances, when better understood,
might provide new and powerful diagnostics of the physical processes
underpinning stellar coronae.  The emerging patterns of coronal
abundance anomalies are telling us something about the dynamical 
structure and heating of coronal plasma; the challenge is to learn 
to read these patterns. Aiming toward this goal, one question that
arises is that of fine tuning of abundance patterns: how do the abundance 
anomalies in the late-type giants compare with those of solar-like and 
other active stars?

In low-mass (M$<$1.5\,M$_\odot$) solar-type stars the magnetic
activity is thought to derive from a dynamo powered by convection and
driven by stellar spin \citep[e.g.][]{Parker70}. Thus extremes in activity
and rotation are closely related.  Moderate-mass giants
(M$\sim$3.0\,M$_\odot$), with A and late B progenitors on the MS
having no outer convection zone, cross into the cool half of the H-R
diagram during their post-main-sequence phase as yellow giants. The
magnetic activity of these stars experiences several stages as they
evolve: the Hertzsprung-gap phase of relatively rapid rotation
\citep{Simon89}; the ``rapid braking zone'' \citep{Gray89} and { the
red giant branch (RGB) phase \citep{Ayres83}. The latter includes the
ascent of the RGB, the helium ignition at its tip, and the return to
the base of the RGB as core helium burning (``clump'') stars. The
rejuvenation of magnetic activity of more evolved ``clump'' giants is
not well understood. Infusion of angular momentum at the base of the
RGB \citep{Simon89} and the rejuvenation of the dynamo by planet/brown
dwarf accretion \citep{Siess99} have been suggested as possible
causes}. Coronal characteristics and X-ray emission, triggered by the
onset of efficient convection occurring while passing through the
Hertzprung-gap, may change in response to evolutionary changes of the
rotation and internal structure of the star (e.g.; deepening of the
convection zone, rapid expansion of the stellar radius and photosphere
cool down). Yellow giants are particularly interesting due to the
relatively short time scales in which internal structure changes which
might affect the corona and the coronal abundances take place
\citep{Ayres98,Gondoin99}.

In this paper, using {\it Chandra} High Energy Transmission Grating
spectrograph (HETGS) observations, we present a comparative analysis
of the coronal X-ray spectra of the active late-type giants $\beta$
Cet, 31 Com and $\mu$~Vel with particular emphasis on their
abundances.  We first describe the three stars and briefly review
earlier work (\S\ref{s:stars}), then in \S\ref{s:obs} we report on the
{\it Chandra} observations and data reduction.  The methods used for a
differential emission measure analysis together with results obtained
are shown in \S\ref{s:anal}. In \S\ref{s:results} we discuss our
results on the coronal abundances and temperature structure and report
our conclusions in \S\ref{s:concl}.

\section{Program Stars}
\label{s:stars}

Our sample represents different giant evolutionary stages: the
Hertzsprung gap (31 Com), the rapid braking zone ($\mu$ Vel) and the
core helium burning ``clump'' phase ($\beta$ Cet). $\beta$ Cet, 31 Com
and $\mu$ Vel are among the more active single G-type giant stars in
the solar neighborhood, lying about two orders of magnitude brighter
in X-rays than the average value for single G-giants
\citep[L$_x$=28.7, ][]{Maggio90}.

\subsection{$\beta$ Cet}

The ``clump'' giant $\beta$ Cet (HD 4128, HR 188) is a well-studied
nearby G9.5\,III star \citep[29\,pc, ][]{Perryman97}. Its main
physical parameters are shown in Table~\ref{t:par}. { Several works
have shown that $\beta$ Cet has a photosphere metal-rich relative to solar (see Table~\ref{t:photabun}). The
enhancement of the C/N abundance ratio \citep[C/N=+0.4, ][]{Lambert81}
clearly shows that $\beta$ Cet has undergone first dredge-up, placing
it, based on the evolutionary time-scale arguments, in the He-burning
clump \citep{Ayres95}}. $\beta$ Cet is the brightest FUV and X-ray
source among the single clump giants. \citet{Smith79} called attention
to the abnormally large photospheric macroturbulence velocity
(4.2$\pm$0.2 km\,s$^{-1}$), and suggested that this could be related
to the UV and X-ray emission if such motions are dissipated as heat by
shocks in the chromosphere and corona. \citet{Ayres01} reported a
34-day EUVE pointing during which a series of coronal flares were
observed. Those long-duration outbursts (several days) are like the
extremes seen on RS CVn type binaries
\citep{Foing94,Garcia-Alvarez03}. An earlier 6-day observation showed
a flat light curve \citep{Ayres94}.  \citet{DrakeS94}, based on ASCA
observations, estimated the coronal abundance of a number of elements
obtaining an atmosphere Mg-rich and S, Fe, O, Ne and N-poor relative
to solar photospheric values. Similar results were reported by
\citet{Maggio98} based on SAX observations.

\subsection{31 Com}

31 Com (HD 111812, HR 4883) is a G0\,III rapidly rotating (P$<7\fd2$)
star 94\,pc away \citep{Perryman97}. It has the bluest color
(B-V=0.67) among the single G giants observed by {\it Einstein
Observatory} \citep{Maggio90}. 31 Com has evolved out of the MS and is
probably crossing the Hertzprung-gap, as suggested by its location in
the H-R diagram and by the high abundance of lithium \citep[log
N(Li)=2.7, ][]{Mallik03}. In this phase 31 Com has developed a
convective subphotospheric layer and dynamo. Metallicity has been
examinated by \citet{Gustafsson74} and \citet{Taylor99}, revealing a
slightly metal-poor photosphere relative to solar (see
Table~\ref{t:photabun}). 31 Com was noted to have moderately weak
\ion{Ca}{2} H\&K lines with reversals
\citep{Wilson66}. \citet{Strassmeier94} reported photospheric line
profile deformations most likely caused by cool
starspots. \citet{Ayres98} suggested rotational modulation of highly
extended chromospheric structures as an explanation of secular changes
in the Ly$\alpha$ profile.

31 Com has been observed several times in X-rays: Einstein
\citep{Maggio90}, ROSAT \citep{Pizzolato00}, ASCA \citep{Ueda01},
XMM-Newton \citep{Scelsi04}. It shows evidence for excess broadening
in {Fe}~{XXI} \citep{Ayres03} which would imply, if the excess
broadening is rotational, that the corona is highly extended, contrary
to the compact structures suggested by recent density estimates in a
number of active coronal sources. \citet{Scelsi04} determined the
emission measure distribution and coronal abundances for 31 Com based
on XMM-Newton observations (RGS and EPIC spectra). They suggested a
corona dominated by isothermal magnetic loops.

\subsection{$\mu$ Vel}

$\mu$ Vel (HD 93497, HR 4216) first appeared in the literature more
than 100 years ago \citep{See1897}, though detailed studies of the
star have been undertaken.  It is a close (2'' apart) visual double
system (G5\,III+dF; 36\,pc away \citet{Perryman97}) consisting of a
late-type giant and a fainter companion.  { \citet{Ayres99} concluded
that the companion is a single mid-F dwarf thought to be relatively
inactive compared with the primary, and thus it should not contribute
significantly to the observed X-ray emission.  We return to this in
\S\ref{s:anal}.}  \citet{Ayres99} reported, based on a deep pointing by
EUVE, a large flare on $\mu$ Vel with a long 1.5 day decay.

\citet{Burnashev83} suggested $\mu$ Vel has a slightly metal-poor
photosphere relative to solar \citep{Holweger79}, [Fe/H]=-0.12.  Based on comparison with
the evolutionary tracks of \citet{Schaller92}, \citet{Ayres99} derived
an age for $\mu$ Vel of $\sim$350\,Myr and classified it as being in
the ``rapid braking zone'' discussed by \citet{Gray89} that begins at
spectral type G0-G3.  Here the {C}~{IV} activity is seen to drop
simultaneously with a decline in rotational velocity.  The properties
of $\mu$ Vel are also fairly similar to those of the X-ray bright
clump giants $\beta$~Ceti and Capella~Aa, including a low lithium
abundance \citep[log N(Li)=0.4][]{Mallik03}.  While it lies very
slightly to the blue of the \citet{Schaller92} core He-burning clump,
the lack of surface C and N abundance measurements precludes a
definitive classification of its exact evolutionary state.  For the
purposes of our study here, however, this exact classification is not
critical and we tentatively adopt the rapid braking zone
classification of \citet{Ayres99}.

\section{Observations}
\label{s:obs}

The {\it{Chandra}} HETGS observation of $\beta$ Cet, 31 Com and $\mu$
Vel were carried out in 2001 using the AXAF CCD Imaging Spectrometer
(ACIS-S) in conjunction with the High-Energy Transmission Grating
Spectrometer (HETGS). All the observations employed the detector in
its standard instrument configuration. The observations are summarized
in Table~\ref{t:obs}. { Note that two observations 3 months apart were
obtained for $\mu$ Vel.}

\begin{figure}
\plotone{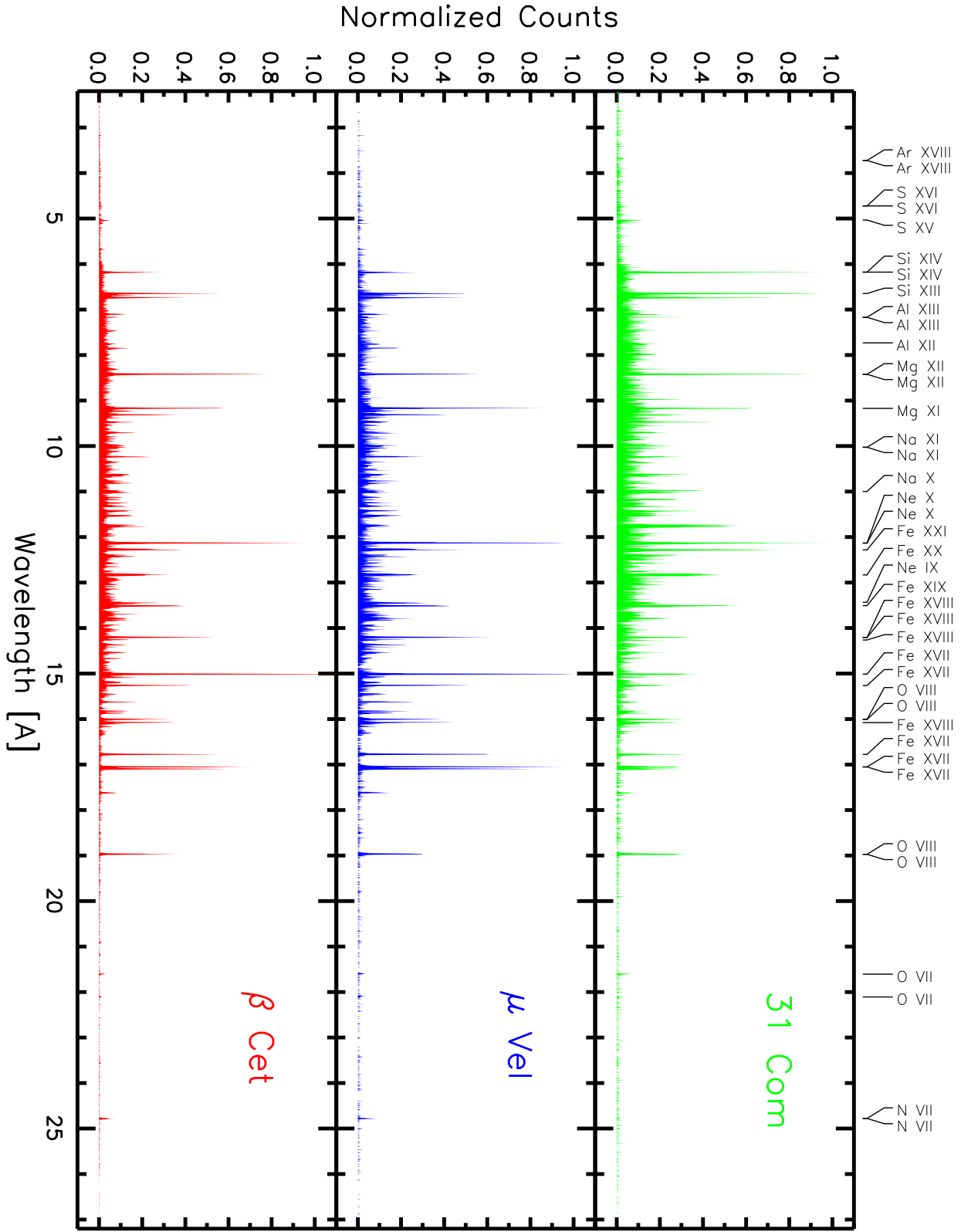}
\caption{{\it Chandra} X-ray spectra of $\beta$ Cet, 31 Com, $\mu$ Vel. The strongest lines over the observed wavelength range are identified.}
\label{f:sp}
\end{figure}

Fig.~\ref{f:sp} shows the {\it Chandra} X-ray spectra of $\beta$ Cet,
31 Com and $\mu$ Vel in the wavelength range 3-27~\AA\ which contains
prominent lines of Ne, N, O, S, Si, Fe, Mg, Al and Na. The strongest
lines are identified. The spectra of $\beta$ Cet and $\mu$ Vel show a
remarkable similarity of lines, both in terms of which lines are
prominent, from H- and He-like ions and the broad range of charge
states of Fe, and in their relative intensities. The spectrum of 31
Com, although similar to the other two stars, shows stronger O and Fe
lines. We also observe that in $\beta$ Cet and $\mu$ Vel the
\ion{Fe}{17} resonance line at 15.01 \AA\ is strongest, in 31 Com
\ion{Ne}{10} Ly$\alpha$ is strongest.

\section{Analysis}
\label{s:anal}

Pipeline-processed (CXC software version 6.3.1) photon event lists
were reduced using the CIAO software package version 3.0, and were
analyzed using the IDL\footnote{Interactive Data Language, Research
Systems Inc.}-based PINTofALE\footnote{Available from http://hea-www.harvard.edu/PINTofALE} 
software suite \citep{Kashyap00}. The analysis we have performed consisted
of line identification and fitting, reconstruction of the plasma
emission measure distribution including allowance for blending of the
diagnostic lines used, and finally, determination of the element
abundances.

\subsection{Photometry}

\begin{figure}
\plotone{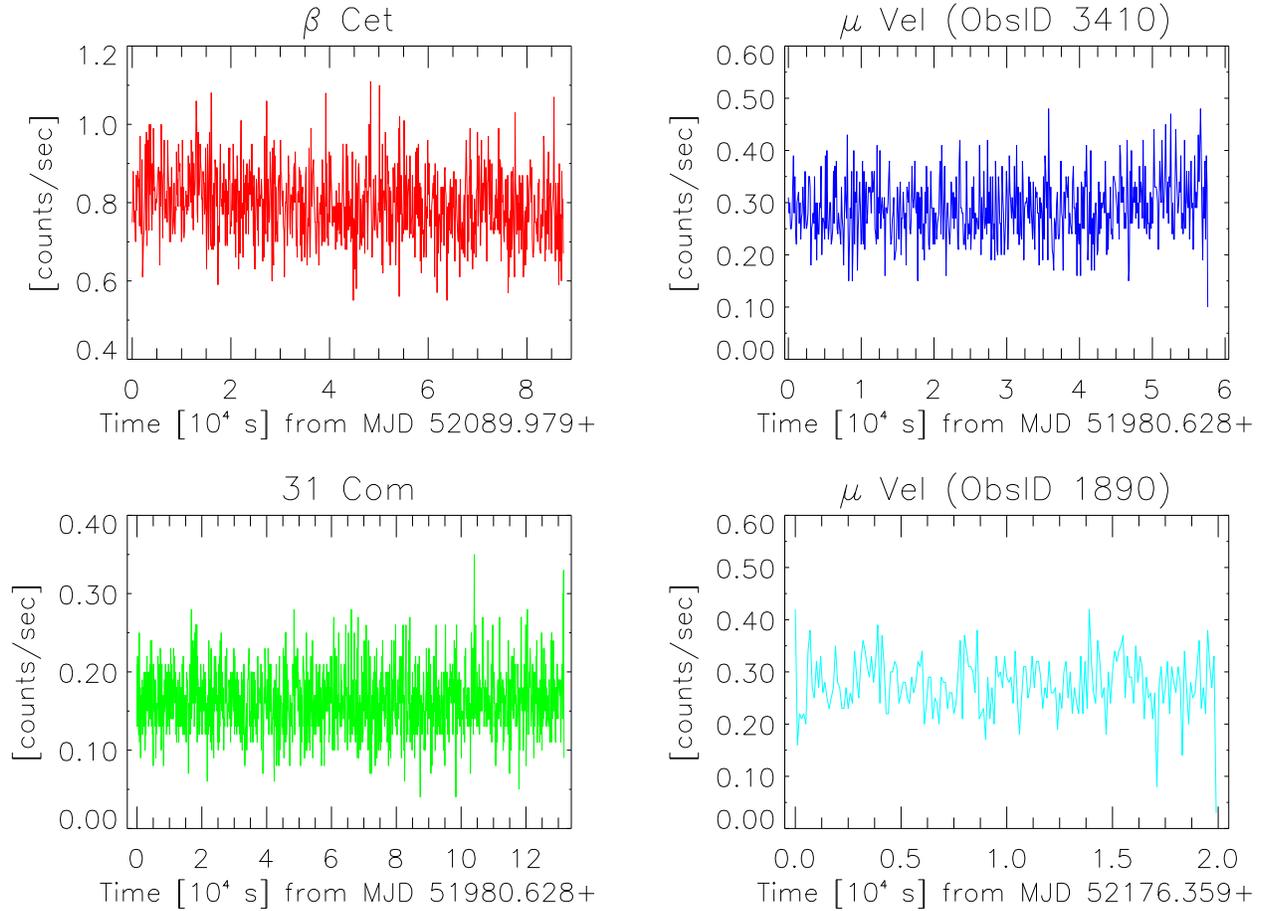}
\caption{{\it Chandra} X-ray light curves of $\beta$ Cet, 31 Com, $\mu$ Vel (ObsID 3410 and ObsID 1890). The light curves are binned at 100s intervals. All objects were relatively quiescent, showing no large flare events.}
\label{f:lc}
\end{figure}

Before commencing spectral analysis, we first checked for flare
activity that could affect not only the shape of the differential
emission measures (DEMs) but might also be accompanied by detectable
changes in the coronal abundances of the plasma that dominate
disk-averaged spectra \citep[see,
e.g.][]{Ottmann96,Mewe97,Ortolani98,Favata99,Favata00,Maggio00,Guedel01}.
Light curves for $\beta$ Cet, 31 Com and $\mu$ Vel observations (including Obs ID1890 and Obs ID3410 hereafter) were derived
excluding the 0th order, which could be affected by pileup, and instead considered just the dispersed photons.  Events were
then binned at 100s intervals. Fig.~\ref{f:lc} shows the light curves for all the observations; note that all the light curves are relatively flat and
devoid of significant flare activity. { The companion of $\mu$ Vel is just discernible in the 0th order X-ray image 
as a slight elongation of the wing PSF.  While the X-ray spectrum of 
this mid-F dwarf cannot be separated from that of $\mu$ Vel, it contributes 
only $\sim 1$\%\ of the total counts and so its influence in the 
analysis will be completely negligible.}

We conclude that $\beta$ Cet, 31 Com and $\mu$ Vel observations are
representative of the stars during times of quiescence, and therefore
treat the observations in their entirety for the remainder of the
analysis.  


\subsection{Spectroscopy}

The emission measure distribution and abundance analysis employed the
CHIANTI database version 4.2 \citep{Dere01} and the ionization balance of
\citet{Mazzota98}, as implemented in the PINTofALE software package
\citep{Kashyap00}. Spectral line fluxes for $\beta$ Cet, 31 Com and $\mu$ Vel were measured by
fitting modified Lorentzian or Moffat (``$\beta$ profile'') functions of the form:
\begin{equation}
f(\lambda)=a/(1+(\frac{\lambda-\lambda_0}{\Gamma})^2)^\beta
\end{equation}
where $a$ is the amplitude and $\Gamma$ a characteristic line width.
For a value of $\beta=2.4$, it has been found that this function
represents the {\it{Chandra}}
transmission grating instrumental profile to within photon counting
statistics for lines of the order of a few
1000 counts or less \citep{Drake04b}. In the
case of blends, as in \ion{Ne}{9}-\ion{Fe}{19} { in which both spectral lines are blended}, we have performed
multi-component fitting { in order to obtain the flux of each line}.

{ While strong lines can be easily identified, the sum
of weak lines, each of which can be undetectable, can produce a
``pseudo-continuum''. The true continuum level was set using the
spectral regions 2.4-3.4, 5.3-6.3 and 20.4-21.4~\AA, which we judged
to be essentially ``line-free'', based on both visual inspection and
the examination of radiative loss models.  These continuum points were
then used to normalize model continua computed using a test
Differential Emission Measure taken from an analysis of a typical 
active star (AU~Mic in this case, \citet{Drake05}).  In
principle, the fluxes should be re-measured once the final DEM has
been determined; in practise, there was little difference in
the {\em shapes} of trial and final continuum models so that this
additional iterative step was not required.}

Table~\ref{t:flx} shows the measured fluxes of the
emission lines identified and used in this analysis. We have computed
correction factors (CF) for line blends. { The procedure identifies which lines, within the CHIANTI emission line database, are possible contaminants
for each analysed line. The contaminants must lie within two sigma of the line center and have predicted intensity greater than a user-defined threshold. Gaussian profiles are used,for simplicity of computation, as an
approximation to the true profiles of the contaminants. The differences compared with the Moffat profiles, used to measure the spectral line fluxes, are not significant.
} The CF are defined as follows:
\begin{equation}
CF = {I(\lambda_i) \over I(\lambda_i) + \sum_{j \neq i} I(\lambda_j)}
\end{equation}
where I($\lambda_i$) = intensity of the line being contaminated by
lines $j \neq i$. CF
for line blends for the studied lines range from 0.73 to
0.99. { The measured fluxes are then multiplied by the CF in order to obtain the ``true'' flux of each line.}

\subsection{Differential Emission Measure}
\label{s:dems}

Our basic set of diagnostics
comprises the H-like/He-like
resonance line flux ratios for the elements O, Ne, Mg, and Si,
line ratios involving Fe~XVII, Fe~XVIII and Fe~XXI
resonance lines, and measurements of the continuum flux at points in
the spectrum that are essentially free of lines (the same regions
noted above with regard to spectral line intensity measurement). 
This set of lines include the brightest lines in stellar coronal spectra and
are easily measured in essentially all well-exposed {\em Chandra}
grating observations of stellar coronae, such that star-to-star
variations in diagnostic lines used can be avoided. 

While atomic data for the H-like and He-like resonance lines used here
are expected to be reliable this is not necessary the case for the Fe
lines, for which both ionization balance and level population
computations are much more complex.  \citet{Doron02} and \citet{Gu03}
reported new Fe L-shell calculations indicating that rate coefficients
for n$\rightarrow$2 (3$<$n$<$5) transitions originating from 3s and 3p
upper levels may differ from earlier calculations up to 50\%.  They
emphasised the importance of including in the level populations the
indirect processes of radiative recombination, dielectronic
recombination (DR), and resonance excitation involving neighbouring
charge states.  However, the \ion{Fe}{17} 15.01\AA\ and \ion{Fe}{18}
14.21\AA\ lines used here arise from $3d$ upper levels which the
calculations indicate are not significantly affected by these
processes.  \citet{Gu03} also found the effects of indirect processes
for \ion{Fe}{20}-\ion{Fe}{25} lines to be smaller than 10\%.  While
these calculations provide additional support for our choice of Fe
lines and add confidence to the propriety of the atomic data,
demonstrable problems still remain in the case of \ion{Fe}{18}.
\citet{Desai05} found good agreement between the predicted fluxes for
the \ion{Fe}{18} lines at 14.21\AA\ and 14.26\AA\ obtained from the
CHIANTI and Flexible Atomic Code \citep{Gu03b}, which includes the DR
processes.  However, the fluxes observed in {\it Chandra} grating
spectra of Capella are $\sim$50\% higher than expected based on
normalisation to the EUV $2s$-$2p$ transitions, with the likely
culprit being the excitation rates.  Such errors in the \ion{Fe}{18}
atomic data might introduce some additional structure into the
best-fit DEM solutions, though this will be tempered by the
statistically more significant information from the H-like and He-like
ions of Mg and Si.  The abundances themselves are quite insensitive to
small deviations in the DEM and we do not expect the possibly errors
in the \ion{Fe}{18} atomic data to have a significant impact.

In order to obtain the {\em differential emission measure} (DEM) we have
performed a Markov-Chain Monte-Carlo analysis using a Metropolis algorithm
(MCMC[M]) on the set of supplied line flux ratios
\citep{Kashyap98}.  Advantages of this method include the ability to
estimate uncertainties and the
avoidance of unnecessary smoothing constraints. The DEM is a measure of the amount of
emitting power---correlated with the amount of emitting material---as
a function of temperature in the coronal plasma. It is formally
defined as: 
\begin{equation}
\Phi(T)={n_e^2}\frac{dV(T)}{dlogT}
\label{e:dem}
\end{equation} 
A given line flux depends on both temperature and
on the abundance of the element in question.  The ratio of two emission lines from ions of the
{\em same} element is independent of the abundance of the chosen
element.  We have therefore devised a method that uses line {\em
ratios} instead of line fluxes directly \citep[][]{Garcia-Alvarez05} (A similar method was developed 
independently by \citet{Schmitt04}). The MCMC[M] method yields the emission
measure distribution over a pre-selected temperature grid, where the
DEM is defined for each T bin. 
In our case, a set of temperatures $T_n$, with
$\Delta\,\log\,T[K]$=0.1 and ranging from $\log\,T$=6.2 to
$\log\,T[K]$=7.5, define the $\Phi(T_n)$. The derived $\Phi(T_n)$ is
only reliable over a certain temperature range if we have enough lines
with contribution function G($T_{max}$)$\sim$
G($T_{n}$).

Based on the lines we use in our analysis we are
able to obtain a well-constrained DEM(T$_n$) between $\log\,T[K]$=6.2 (coolest peak formation temperature given by the resonance line \ion{O}{7} is $\log\,T[K]$=6.3)
and $\log\,T[K]$=7.5 (hottest
peak formation temperature given by the resonance line \ion{S}{16} is $\log\,T[K]$=7.4); larger uncertainties are obtained outside that
range. The H-like and He-like lines constrain the 
structure of the DEM in the upper and lower temperature range, while 
the iron ions give information about the structure of the DEMs in the 
intermediate range. The MCMC[M] code returns the DEM, among any user-defined number
of possible DEMs that are 
generated based on the MCMC method, that best reproduces the observed
fluxes.  
As noted above, one product of our MCMC[M] method is the statistical
uncertainties in $\Phi(T)$ evaluated at the 68\% confidence level. One
has to bear in mind that 
the error bars for consecutive bins in DEMs are not independent,
owing to the inevitable cross-talk that arises as a result of the line
contribution functions themselves stretching over several bins.
While our DEM method imposes a degree of local smoothing, dictated by
the number of diagnostic lines with significant emission in each
temperature bin \citep[see][ for details]{Kashyap98}, this smoothing is
not always sufficient to attenuate ``high frequency oscillations'' in
the solution---the artifact whereby overly high values of the DEM in a
given temperature bin can be compensated for by a commensurately 
lower value in adjacent bins. { In principle, there is no limit to the scale of the high-frequency
oscillations. In practice, the binning inherent in the temperature
grid serves as a de-factor smoothing scale that eliminates all
oscillations unresolved by the grid.} Our final DEMs are corrected for the distances of the stars. 

\subsection{Abundances}

{ In comparing the coronal abundance results with underlying photospheric 
values it is tempting to adopt a photospheric mixture different to that 
of the Sun to accomodate small deviations from such a mixture that might 
have been found in the different photospheric studies. However, as 
discussed in \S\ref{s:stars}, of our sample only the photospheric 
composition of $\beta$~Ceti has been studied in any detail. Close 
examination of the results listed in Table~\ref{t:photabun} (which have 
been scaled relative to the abundance mixture of \citet{Asplund04}), however, 
reveals very little evidence for a significant deviation from a solar 
mixture.  Exceptions are, possibly, Si, which appears slightly 
over-abundant compared to Fe in the \citet{Ottmann98} study, with 
an indication of a similar trend in the earlier work of 
\citet{McWilliam90} and \citet{Luck95}.  Na also appears slightly over-abundant in 
the studies of \citet{Kovacs83}, \citet{Luck95}, and Drake 
(unpublished).  The latter might be related to the Ne-Na cycle which can 
enhance the Na abundance in the post-dredge-up envelopes of giants stars 
\citep[e.g.][]{Denisenkov87,Drake94,Langer95,Weiss00}. 
However, as becomes clear in the coronal abundances, there is no 
evidence for significantly enhanced Si, and we have only upper limits 
for the fluxes of the available Na lines in our spectra.  We therefore 
adopt the mixture of \citet{Asplund04} for comparing all coronal 
abundanc results. \citet{Asplund04} revision of the solar chemical composition 
shows a downward by 25-35\%\ of the abundances of light elements such as 
C, N, O and Ne compare with values from earlies studies \citep[e.g.][]{Grevesse98}. 
The use of these new values would increase the derived stellar coronal abundances of 
light elements relative to those obtained by using the solar chemical composition from earlies 
studies. In order words, the observed trends in the FIP effect and in the inverse-FIP effect will be slighlty 
shallower and slightly steeper respectively if one uses the new solar chemical composition reported by \citet{Asplund04}.

One exception to the \citet{Asplund04} mixture might be the 
element Ne. \citet{Drake05b} have recently found Ne/O to be 
enhanced by an average factor of 2.7 compared to the 
\citet{Asplund04} ratio in the coronae of approximately 20 nearby 
stars. Based on this results, we have assumed a Ne abundance revised upward by 0.43\,dex
with respect to \citet{Asplund04}. In the discussion that follows, we therefore also examine the 
effects of a higher photospheric Ne abundance.}

Once the DEM has been established, we can evaluate the abundances of
any elements for which we have lines with measured fluxes.  We derive
values for the coronal abundances of the elements Ne, O, S, Si, Fe,
Mg, Al and Na, though only upper limits were obtained for the
abundances of Na and Al, owing to the lack of signal in their
respective line features (see Table~\ref{t:abund1} and
Table~\ref{t:abund2}). If there is more than one line for an element,
then the abundance is computed as the weighted average of the
abundance determinations from each of the individual lines.

We also used the temperature-insensitive abundance ratio diagnostics
of \cite{Drake05} as an additional check on our values obtained using
the DEM.  These are ratios formed by combining two sets of lines of
two different elements, constructed such that the combined emissivity
curves of each set have essentially the same temperature
dependence. The resulting ratio of measured line fluxes then yields
directly the ratio of the abundances of the relevant elements,
independent from the atmospheric temperature structure. A similar,
though less sophisticated method, has been previously used in solar
coronal observations \citep[e.g.][]{Feldman92,Feldman00} and in
stellar EUVE observations \citep[][]{Drake95,Drake97}.

\section{Results and Discussion}
\label{s:results}

\begin{figure}
\plotone{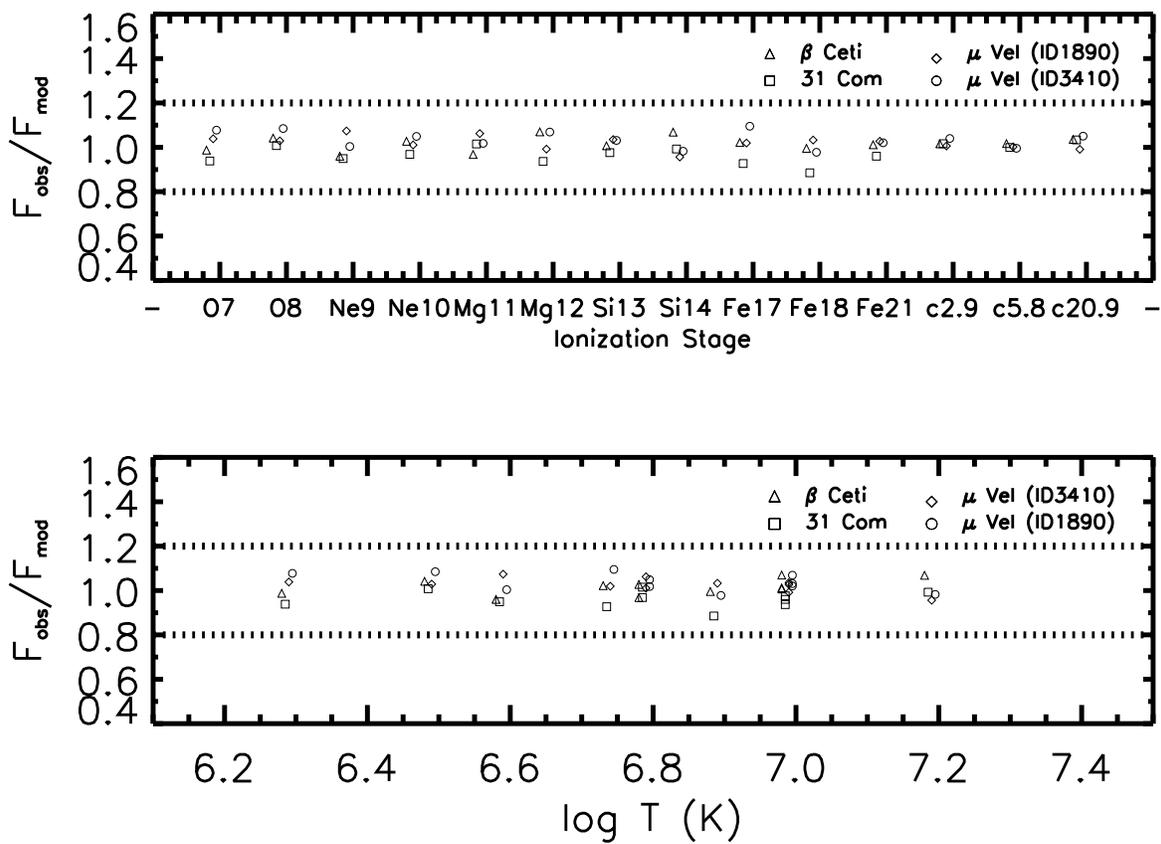}
\caption{Comparison of observed and modelled line fluxes vs ionic
species (upper panel) and vs $T_{max}$ (lower panel) for $\beta$ Cet, 31 Com, $\mu$ Vel (ObsID 3410 and ObsID 1890). { The ``line-free'' continuum spectral regions at 2.4-3.4, 5.3-6.3 and 20.4-21.4 are labelled as c2.9, c5.8 and c20.9 respectively.}}
\label{f:obs_pre}
\end{figure}

In order to verify the propriety of our DEM and abundance techniques
we have compared observed and modeled line fluxes vs ionic species in
Fig~\ref{f:obs_pre} (upper panel) and vs $T_{max}$ (lower panel) for
$\beta$ Cet, 31 Com and $\mu$ Vel (ID1890 and ID3410) and for the
final DEMs and abundances. All the predicted line fluxes based on our
final models are within 10~\%\ or so of the observed values.

\begin{figure}
\plotone{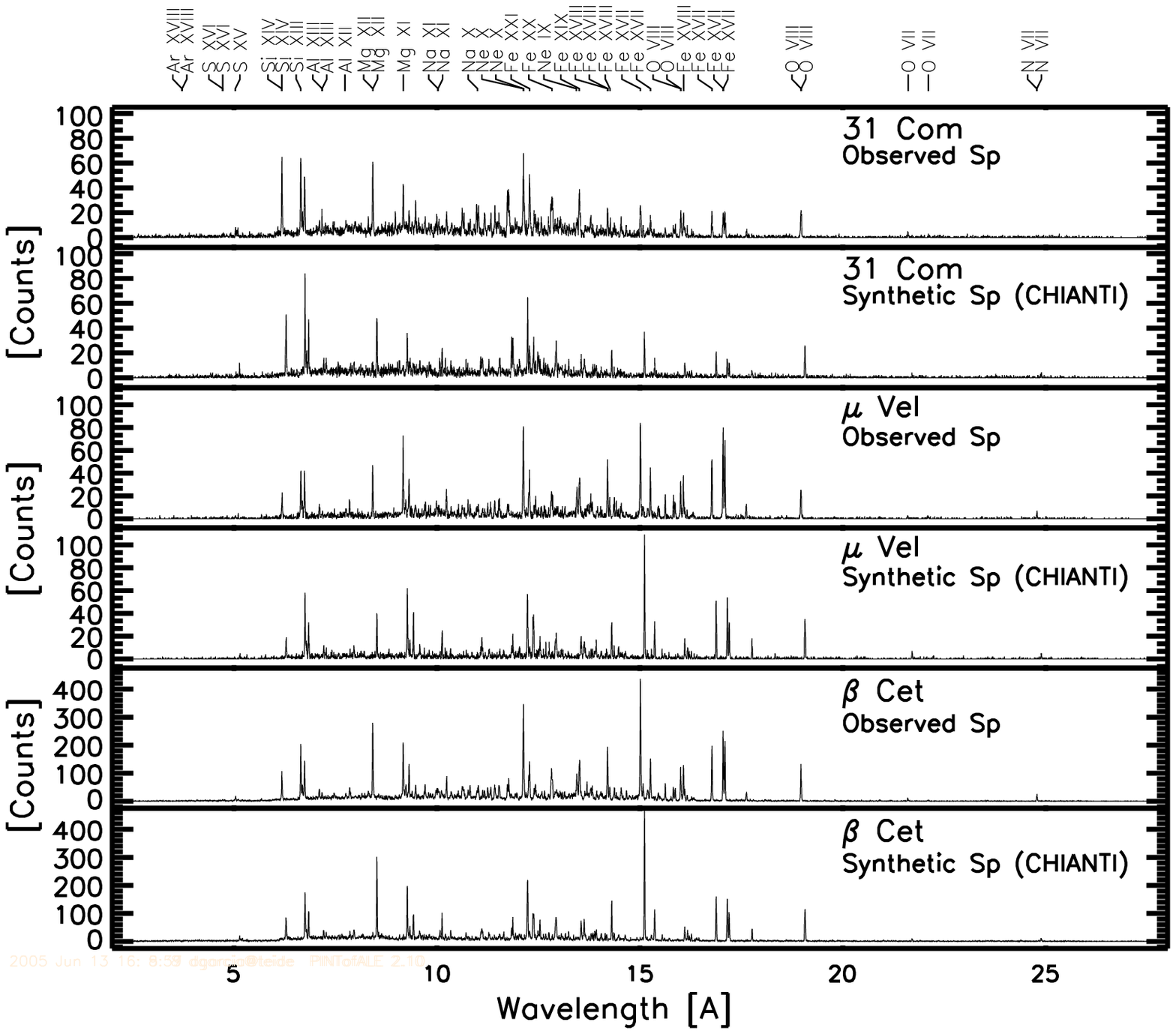}
\vspace{1cm}
\caption{Comparison of predicted
and observed HETG+ACIS-S spectra of program stars. The predicted spectra
are computed using the reconstructed DEMs (see \S\ref{s:structure}).  The predicted
spectrum for $\mu$\,Vel is a sum of the spectra computed using the two
DEMs for the two ObsIDs.  The predicted spectra are offset for
clarity.}
\label{f:synsp_abdor}
\end{figure}

Fig~\ref{f:synsp_abdor} shows the observed and synthetic spectra of
$\beta$ Cet, 31 Com and $\mu$ Vel. For these spectra, we have used an
electron density of n$_{e}$= 1.8\,10$^{12} cm^{-3}$ for $\beta$~Cet,
n$_{e}$=3.2\,10$^{12} cm^{-3}$ for 31~Com and n$_{e}$=1.0\,10$^{12}
cm^{-3}$ for $\mu$~Vel derived by \citet{Testa04}, and an interstellar
medium column density of n$_{H}$=1\,10$^{18} cm^{-2}$. { Although the
n$_{H}$ for $\mu$ Vel could be underestimated the exact value for the
ISM column makes no practical difference to the synthetic spectra
below 50 \AA}. All the synthetic spectra showing good qualitative
agreement with observations for the three stars.

\subsection{Temperature Structure}
\label{s:structure}

\begin{figure}
\plotone{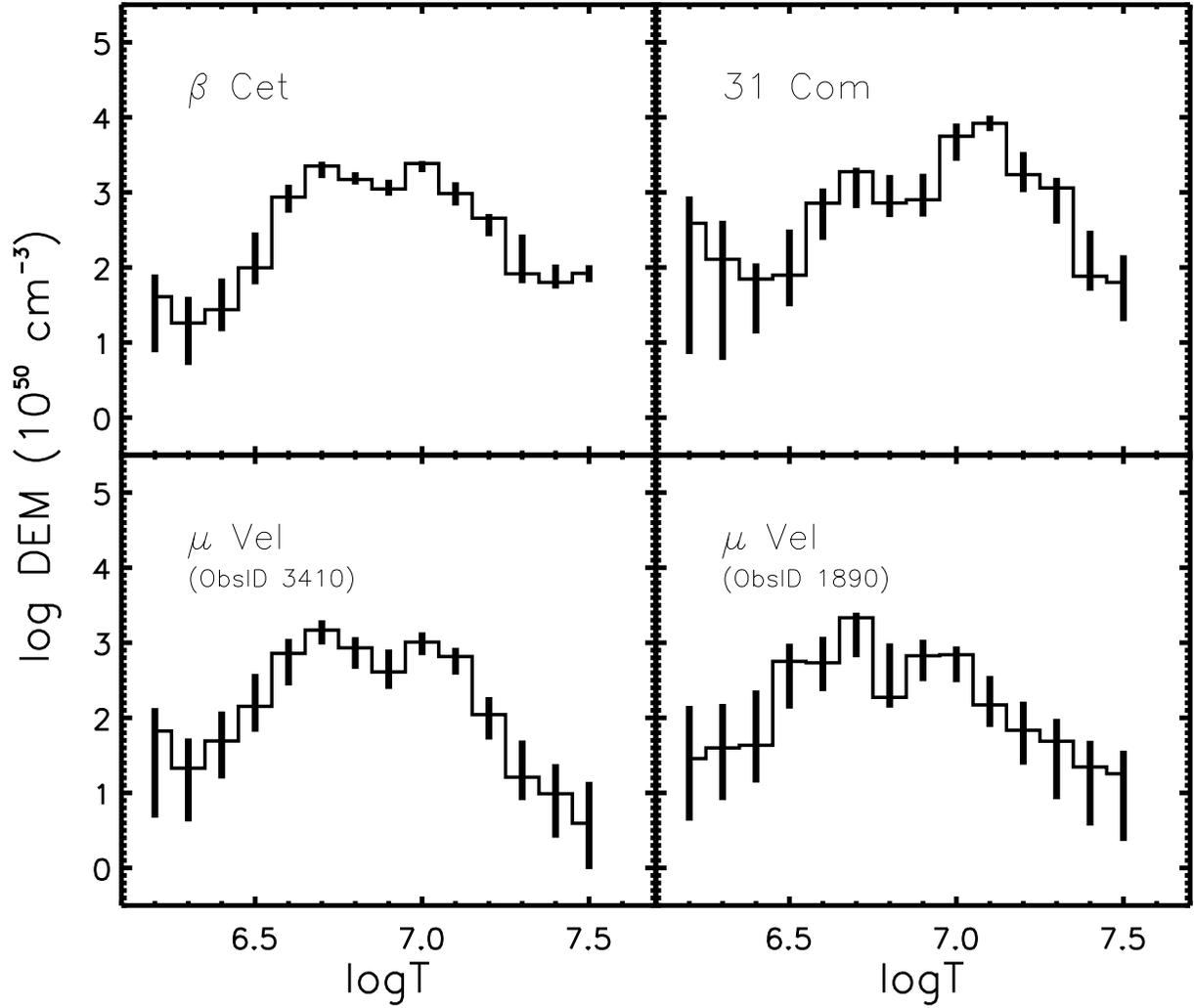}
\caption{DEMs obtained for $\beta$ Cet, 31 Com, $\mu$ Vel (ObsID 3410 and ObsID 1890) by running a
MCMC[M] reconstruction code on a set of lines of
H-like, He-like and highly ionized Fe line fluxes (O, Ne, Mg, Si, \ion{Fe}{17}, \ion{Fe}{18} and \ion{Fe}{21}).}
\label{f:dems}
\end{figure}

Fig.~\ref{f:dems} shows the reconstructed DEMs for $\beta$ Cet, 31 Com
and $\mu$ Vel. The DEMs for $\beta$ Cet and $\mu$ Vel are all quite
similar in overall shape and normalisation although the former is 2-3
times brighter.  Both exhibit a broad peak around $\log\,T$[K]
$\sim$6.8$\pm$0.2, while 31~Com shows a smooth increase reaching a
peak at $\log\,T$[K]$\sim$7.1. The DEMs for the three stars show
little evidence for substantial amount of plasma at temperatures
higher than $\log\,T$[K]=7.3. Nevertheless, 31 Com seems to have more
plasma at higher temperatures compared with $\beta$ Cet and $\mu$
Vel. It also has a higher X-ray luminosity (see Table~\ref{t:par}).

We note the similarity in $\beta$ Cet, 31 Com and $\mu$ Vel thermal
structures in the low temperature range ($\log\,T$[K]=6.4-6.7).  For
$\mu$ Vel and $\beta$ Cet we find a perfect match even for lower
temperatures and for higher temperatures up to $\log\,T$[K]=7.2. The
slopes in the derived DEMs on the low temperature side of their maxima
follow very approximately the relation DEM$\propto$T$^{5}$ for $\beta$
Cet and $\mu$ Vel and DEM$\propto$T$^{5/2}$ for 31 Com, which is
steeper than the DEM$\propto$T$^{3/2}$ suggested for quasi-static
coronal loops \citep[e.g.;][]{Craig78,Jordan80,Peres00}.

Our results for 31 Com are comparable with those reported by
\citet{Ayres98} and \citet{Scelsi04} based on {\it EUVE} and
XMM-{\it{Newton}} observations respectively. \citet{Scelsi04} obtained
a single-peak DEM with a sharper increase for the lower
temperatures, $\log\,T$[K]$<$6.8, and smooth decrease for the higher
temperatures, $\log\,T$[K]$>$7.2. The \citet{Scelsi04} line-based DEM
also shows a maximum at $\log\,T$[K]$\sim$7.0$\pm$0.1. \citet{Ayres98}
also derived, based on {\it EUVE} observations, the DEM for $\beta$
Cet. They obtained a thermal structure that peaks at
$\log\,T$[K]$\sim$6.8 showing a smooth decrease thereafter. To our
knowledge, there exist no other DEM analysis of $\mu$ Vel by other
authors with which to compare our temperature structure results.
\subsection{Coronal Abundances}

Our results for the element abundances of $\beta$ Cet, 31 Com and
$\mu$ Vel are summarized in Table~\ref{t:abund1}. We only list
statistical uncertainties here.  The formal error estimate for the
abundances reflects only the uncertainty of the observed fluxes
together with the uncertainty in the estimated DEM.  This is done by
examining the distribution of synthetic fluxes around the best fit
flux, as computed from the collection of DEMs sampled by the MCMC
method.  There are of course other uncertainties in the atomic data
used for the analysis, and in the calibration of the {\it Chandra}
instruments.  Atomic data uncertainties are probably of order 30\%\
for Fe lines \citep[e.g.;][]{Drake95}, but are probably slightly less
than this for H-like and He-like lines. These uncertainties are,
however, systematic in nature and will partially cancel when lines of
different ionization stages and multiplets are used to estimate
abundances.  Uncertainties in instrument calibration are of order
10\%\ and will also be systematic in nature.  The final uncertainties
in our derived abundances are difficult to assess rigorously, but are
likely of order 0.1~dex.

The abundance ratios derived from temperature-insensitive line ratios
are listed in Table~\ref{t:abund2}.  These values are in agreement
with, and provide verification for, the ones derived from the DEMs
(Table~\ref{t:abund1}), with the exception of the ratios involving
Fe. As noted by \citet{Drake05}, however, these ratios are likely to
exhibit larger systematic errors owing to a less optimum coincidence
of the different line contribution functions as a function of
temperature.

\begin{figure}[t]
\plotone{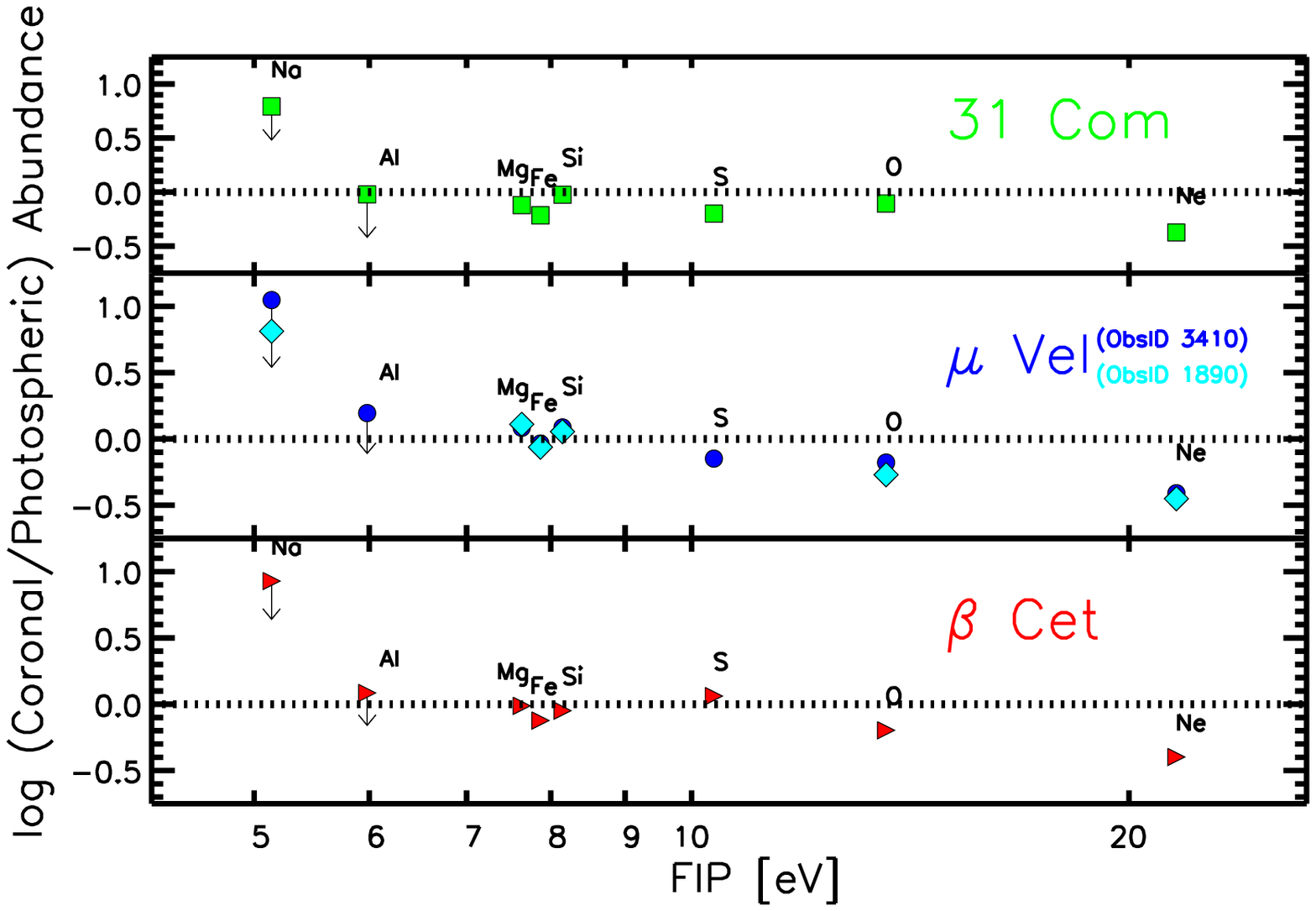}
\caption{\small{Coronal abundances obtained from abundance-insensitive DEM reconstruction for $\beta$ Cet, 31 Com, $\mu$ Vel (ObsID 3410 and ObsID 1890), relative to the abundance mixture of \citet{Asplund04} with Ne from \citet{Drake05b} (see Table~\ref{t:photabun} and discussion in \S5.2). Note: The abundances have been calculated using DEMs derived from set of ions O, Ne, Mg, Si, \ion{Fe}{17}, \ion{Fe}{18} and \ion{Fe}{21} for each object. The errors in the coronal abundances are $\sim$ 0.1 dex.}}
\vspace{-.6cm}
\label{f:abund}
\end{figure}

In Fig.~\ref{f:abund} we have plotted the derived abundances, relative
to the adopted stellar photospheric values (essentially the mixture of
\citet{Asplund04}, with the exception of Ne for which we adopt the
\citet{Drake05} value, as discussed earlier), in order of element FIP,
for $\beta$ Cet, 31 Com and $\mu$ Vel.  Despite almost an order of
magnitude range in rotational velocity among this small sample of
late-type giants they clearly share similarities in coronal
composition. Our analysis shows a slight change in the trend of the
coronal abundance vs FIP from the less evolved 31 Com (very little or
no trend with FIP) to the more evolved $\beta$ Cet and $\mu$ Vel (some
trend with FIP discernible). $\beta$ Cet and $\mu$ Vel show the same
general abundance pattern, characterised by a depletion of high FIP
elements (O, Ne) by a factor of $\sim 2$, and with no evidence for
significant depletions of the low FIP elements (Mg, Fe and Si)
relative to stellar photospheric abundances.  

The FIP effect observed for $\beta$ Cet and $\mu$ Vel is similar to
those seen in other low to intermediate activity stars
\citep{DrakeS94,Drake95,Drake97,Guedel02}.  We emphasise that there is
some uncertainty in the absolute normalisation of the derived
abundances compared with assessed photospheric values, such that we
cannot determine with certainty whether or not the low FIP elements in
$\beta$ Cet and $\mu$ Vel are truly photospheric and the high FIP
elements depleted, or whether the high FIP elements are photospheric
and the low FIP elements are enhanced.  We also note that there is no
definitive information regarding the exact photospheric abundances of
the elements in question, and observed coronal abundance trends should
be treated with some caution.  Nevertheless, the similarity of the
abundance results for $\beta$ Cet and $\mu$ Vel and the general FIP
effect trends seen for other stars compared with the well-established 
solar photospheric baseline does lend some support to the coronal
abundance anomaly interpretation.

\begin{figure}[t]
\plotone{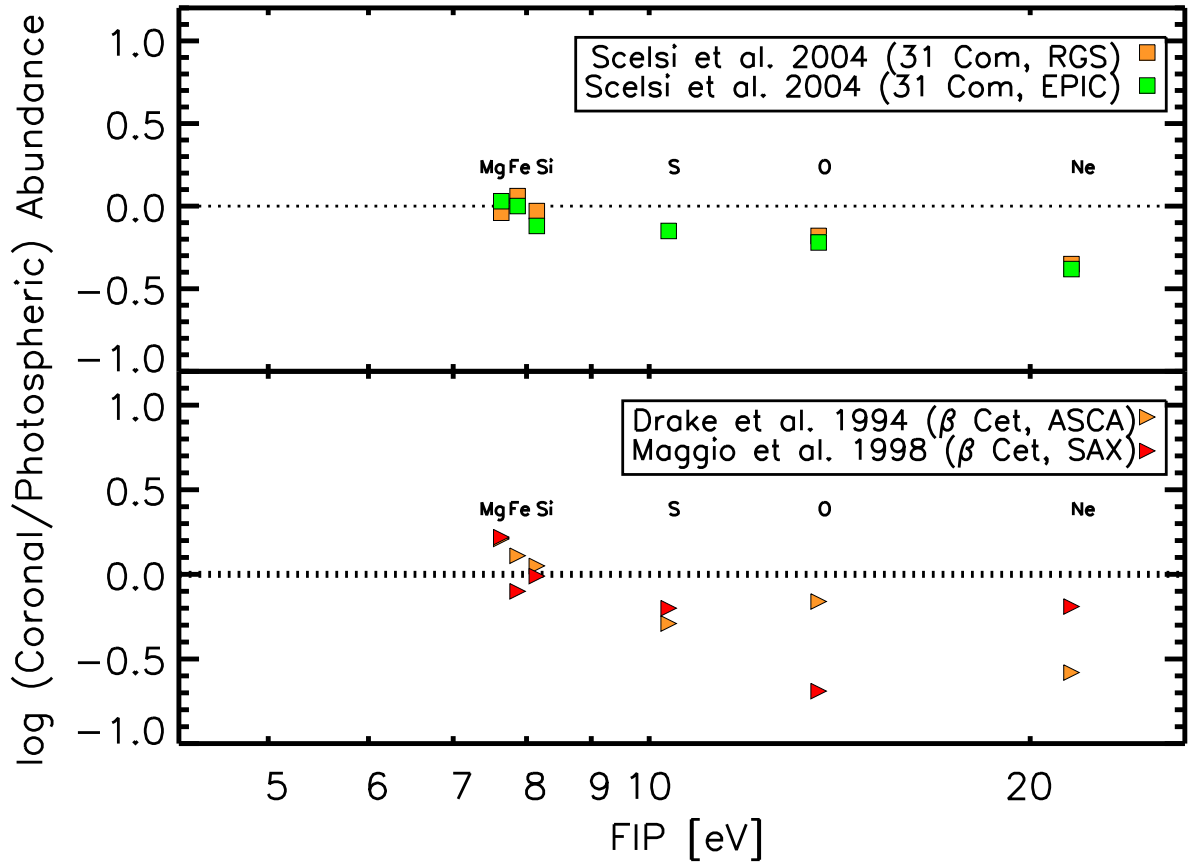}
\caption{\small{Comparison of coronal abundances against FIP for $\beta$ Cet and 31 Com obtained from the literature. The values have been scaled relative to the abundance mixture of \citet{Asplund04} with Ne from \citet{Drake05b} (see discussion in \S5.2)}}
\vspace{-.6cm}
\label{f:abund_liter}
\end{figure}

We also add slight caution in interpreting the absolute Ne abundance.  It
can be seen from comparison of the Ne/O ratios obtained from DEM
modelling and temperature-insensitive line diagnostics
(Table~\ref{t:abund2}) that the former method tends to arrive at a
slightly lower Ne abundance than the latter.  Uncertainties in both
methods are of order 0.1~dex \citep[see also][]{Drake05}, so the
current discrepency is not beyond the bounds of the systematic errors
of the two approaches.  The \citet{Drake05} study suggests Ne is
enhanced with respect to O by a factor of 2.7 as compared to
\citet{Asplund04}: while the temperature-insensitive ratios are in
general agreement with this, our DEM results indicate a number
somewhat lower, and closer to a factor of 1.7 or so.  Adopting a lower
value for the photospheric Ne abundance would raise the Ne points in
Figure~\ref{f:abund} (and in Fig.~\ref{f:abund_liter} described
below).  While it is not possible to state definitively which Ne/O
abundance ratio is the more accurate, it is possible that the DEM
method might slightly overestimate the O abundance relative to that of
Ne by virtue of the DEM being restricted to temperatures $\log T>
6.3$: omission of the cooler material can be compensated for in the
analysis of line fluxes by an increased abundance.  In the case of O,
whose Ne-like and H-like lines are formed at cooler temperatures in
general than those of Ne, this might plausibly lead to a slightly
lower Ne/O ratio.

In Fig.~\ref{f:abund_liter} we plot the abundance results from earlier
studies of $\beta$~Cet and 31~Com \citep{DrakeS94,Maggio98,Scelsi04}.
It is important to note that different authors adopt different values
for ``solar'' photospheric abundances: for the
Fig.~\ref{f:abund_liter} we have scaled the various abundance results
to the \citet{Asplund04} mixture. Despite the fact that the coronal
abundances for $\beta$ Cet were derived from low resolution spectra
there is reasonable agreement with our results in the general trend
observed as a function of FIP for both $\beta$~Cet and 31~Com.

Despite Mg, Fe and Si having very similar FIP, there is a hint in our
results for a small depletion of Fe relative to Mg and Si
by $\sim 0.1$~dex.  This has also been seen in other coronal abundance
studies, as summarised by \citet{Drake03a} who suggested gravitational
settling of the heavier Fe ions as a possible explanation.  However,
the low surface gravity of these giants argues against such a
mechanism---gravitational settling should be negligible in these
stars. Instead, a more likely explanation is systematic error.

It is tempting to interpret the small differences in coronal abundance
between 31~Com and the more evolved giants in terms of trends with
stellar activity. As first suggested by schematically by
\citet{Drake95} and later fleshed out by \citet{Audard03} for active
binaries and \citet[e.g.][]{Telleschi05} for dwarfs, the FIP effect
characterising lower activity stars, including the Sun, is seen to
change to an inverse-FIP effect in more active stars.  Although the
trends observed in our work change in a similar manner, relative to
the rotation of these stars, we speculate instead that the differences
can be attributed primarily to their different evolutionary status and
consequent differences in the properties of their envelopes that are
likely the site of underlying dynamo activity.

These giant stars have A and late B progenitors on the main sequence
which have no outer convection zones. As a late-type giant evolves
from the Herztsprung gap (31 Com) to the ``clump'' ($\beta$ Cet), the
developing convection zone deepens and magnetic braking sets in
\citep[e.g.\ ][]{Gray89}.  Such structural changes might well be
expected to affect dynamo processes and coronal activity.  Indeed,
\citet{Pizzolato00} reported that intermediate-mass stars with
0.5$<$B-V$<$0.8 (31 Com) follow L$_x\sim$(v\,sin\,i)$^2$
\citep{Pallavicini81} while stars redder than B-V=0.8 ($\beta$ Cet and
$\mu$ Vel) rest well above that law. In other words, these two groups
of stars show almost the same X-ray luminosity despite a decrease in
rotational velocity of almost two orders of magnitude. The maintenance
of X-ray luminosity is naturally attributed to the deepening of the
convective zone and possibly to an increase in differential rotation
triggered by magnetic braking.

It is also interesting that large flares have been observed on both
$\beta$~Cet and $\mu$~Vel, yet 31~Com appears to have remained steady
\citep[][and earlier work referenced therein]{Ayres03}.  Similar
differences in gap-clump long-term variability was inferred by
\citet{Johnson02} based on HST STIS observations of Fe~XXI in Capella:
the clump component appears to have varied considerably compared with
earlier observations, yet the gap component was observed at the same
flux level.  

While there are good reasons based on stellar structural arguments to
expect differences in the coronae of gap and clump giants, the actual
mechanisms responsible for the different abundances and temperature
structures remain highly speculative for now.  A recent model based on
ponderomotive forces resulting from Alfv{\' e}n waves
\citep{Laming04} is interesting in this regard.  According to this model,
the FIP effect results from a resonance between the chromospheric
Alfv{\' e}n wave spectrum engendered by turbulence in the convection
zone and coronal loop length that enables the wave ponderomotive force
to either suppress or enhance the movement of ions into the corona.
The FIP effect or its absence therefore depends on the tuning between
the coronal magnetic field and the Alfv{\' e}n wave spectrum.  As we
have mentioned above, the Hertzsprung gap star 31 Com should have a
shallower convection zone and perhaps a different differential
rotation profile with depth compared with $\beta$ Cet and $\mu$ Vel,
which are more evolved.  One might then expect different tuning between
coronal structures and the ambient Alfv{\' e}n wave spectrum.
Qualitatively, this could give rise to the observed abundance
differences.  Detailed calculations of the Alfv{\' e}n wave spectrum
for realistic coronal structures on yellow-red giants will be needed
to confirm or disprove such a model.

\section{Conclusions}
\label{s:concl}

$\beta$ Cet, 31 Com and $\mu$ Vel represent the important stages
through which intermediate mass late-type giants evolve during their
lifetime (the Hetzsprung gap (31 Com), the rapid braking zone ($\mu$
Vel) and the core helium burning ``clump'' phase ($\beta$ Cet)).  As
such, a comparison of their coronal properties provides an
illuminating glimpse of any fundamental underlying differences in
their magnetic dynamos and activity.  Based on an analysis of high
resolution {\it Chandra} X-ray spectra of these stars we draw the
following conclusions.
\begin{enumerate}

\item $\beta$ Cet and $\mu$ Vel show coronal temperature structures
and element abundances that are remarkably similar.  Element
abundances are characterized by a mild FIP-type effect in which the
abundances of low FIP elements are enhanced relative to those of high
FIP elements by a factor of $\sim 2$.  While we cannot rule out this
result as being a fluke of underlying photospheric, rather than
coronal, abundances, the latter scenario is supported by 
similarities with abundance anomalies seen in low-intermediate
activity dwarfs.
 
\item The coronal temperature structure of 31 Com differs from those of
$\beta$ Cet and $\mu$ Vel and exhibits a sharper peak at higher 
temperatures, as seen earlier by \citet{Ayres98} based on EUVE spectra.
Its corona is significantly hotter,
as is evident from comparison of its spectrum with those of $\beta$
Cet and $\mu$ Vel: lines formed at hotter temperatures
are stronger in 31 Com, and it also
has a stronger short wavelength continuum. Element abundances are
characterized by a lack of an obvious FIP effect and appear closer
to photospheric estimates.

 
\item We speculate that structural changes during the evolution of
late-type giants are likely responsible for the small observed
differences in coronal abundances and temperature structure.  In
particular, the size of the convection zone coupled with the rotation
rate seem obvious choices for playing a key role in determining
coronal characteristics.


\end{enumerate}

\acknowledgments

DGA and WB were supported by {\it{Chandra}} grants GO1-2006X and
 GO1-2012X. LL was supported by NASA AISRP contract NAG5-9322; we
 thank this program for providing financial assistance for the
 development of the PINTofALE package. We also thank the CHIANTI
 project for making publicly available the results of their
 substantial effort in assembling atomic data useful for coronal
 plasma analysis.  JJD and VK were supported by NASA contract
 NAS8-39073 to the {\it{Chandra}}.
\clearpage


\newpage\section*{Tables}

\begin{deluxetable}{llrcccccccc}
\tabletypesize{\scriptsize}
\tablecaption{Summary of Stellar Parameters. \label{t:par}}
\tablewidth{0pt}
\tablehead{\colhead{Star} & \colhead{HD}&\colhead{Sp. Type} & \colhead{B-V} & \colhead{D} &  
\colhead{$R_\odot$} & \colhead{$M_\odot$} & \colhead{$g_\odot$}& \colhead{T$_{eff}$} &  \colhead{$v$\,sin$i$} & \colhead{$\log$ L$_x$}\\
\colhead{} & \colhead{}&\colhead{} & \colhead{} & \colhead{[pc]} &  
\colhead{} & \colhead{} & \colhead{}& \colhead{[K]} &  \colhead{[km\,s$^{-1}$]} & \colhead{[erg\,s$^{-1}$]}} 
\startdata
31 Com & 111812& G0\,III\tablenotemark{e} &  0.67\tablenotemark{e} & 94\tablenotemark{b} & 9.3\tablenotemark{e} &2.9\tablenotemark{e} &0.039&5320\tablenotemark{h}&66.5\tablenotemark{f}&30.9\tablenotemark{d} \\
$\mu$ Vel & 93497& G5\,III\tablenotemark{a} &  0.91\tablenotemark{a} & 36\tablenotemark{b} & 13.0\tablenotemark{g} &3.0\tablenotemark{g}&0.016 &4862\tablenotemark{i}&6.4\tablenotemark{f}&30.3\tablenotemark{d} \\
$\beta$ Cet& 4128& G9.5\,III\tablenotemark{a} &  1.02\tablenotemark{a} & 29\tablenotemark{b} & 15.1\tablenotemark{c} &3.0\tablenotemark{c} &0.022 &4840\tablenotemark{h}&3.0\tablenotemark{c}&30.4\tablenotemark{d} \\
\enddata
\tablenotetext{a}{\citet{Makarov03}}
\tablenotetext{b}{\citet{Perryman97}}
\tablenotetext{c}{\citet{Ayres98}}
\tablenotetext{d}{Derived from our {\it{Chandra}} observations (0.5-2.4 KeV).}
\tablenotetext{e}{\citet{Pizzolato00}}
\tablenotetext{f}{\citet{deMedeiros95}}
\tablenotetext{g}{\citet{Ayres99}}
\tablenotetext{h}{\citet{Gondoin99}}
\tablenotetext{i}{\citet{Favata95}}
\end{deluxetable}

\begin{table}
\caption{$\beta$ Cet, 31 Com, $\mu$ Vel photospheric abundances$^{a}$ from literature and adopted here.\label{t:photabun}}
\scriptsize{
\begin{center}
\begin{tabular}{cccccccccccl}
\hline  \hline \\
$\beta$ Cet&&&&&&&&&&& \\\\
{T$_{eff}$}&
{log g}&
{$\xi$}&
{$[$Na/H$]$}&
{$[$Al/H$]$}&
{$[$Mg/H$]$}&
{$[$Fe/H$]$}&
{$[$Si/H$]$}&
{$[$S/H$]$}&
{$[$O/H$]$}&
{$[$Ne/H$]$}&
{References}\\\\
\hline
4\,810 & 2.34 & 1.5 & \nodata&\nodata & \nodata&0.00$\pm$0.15 & \nodata& \nodata&+0.02$\pm$0.16 & \nodata& \citet{Lambert81}\\
4\,800 & 2.50 & 2.3 &+0.63$\pm$0.24 &+0.23 &+0.01$\pm$0.28 &-0.10$\pm$0.16 &+0.04$\pm$0.27 &\nodata &\nodata &\nodata & \citet{Kovacs83}\\
4\,840 & 2.80 & \nodata & \nodata& \nodata&\nodata &+0.16 &\nodata &\nodata &\nodata & \nodata& \citet{Burnashev83} \\
4\,860 & 2.80 & 1.6 &\nodata & \nodata&\nodata &+0.09$\pm$0.08 & \nodata&\nodata &+0.15$\pm$0.19 &\nodata & \citet{Gratton86}\\
4\,800 & 2.40 & 1.7 & \nodata&\nodata & \nodata&+0.26 & \nodata& \nodata& \nodata& \nodata& \citet{Brown89}\\
4\,820 & 2.87 & 2.2 & \nodata&\nodata & \nodata&+0.13$\pm$0.11 & 0.22& \nodata& \nodata& \nodata& \citet{McWilliam90}\\
5\,000 & 2.75 & 1.7 & \nodata&\nodata & \nodata&+0.37$\pm$0.12 & \nodata& \nodata&\nodata& \nodata& \citet{Jones92}\\
4\,750 & 2.65 & 2.4 & +0.52$\pm$0.12&+0.23 & \nodata&+0.20$\pm$0.27 & +0.42$\pm$0.18& \nodata& \nodata& \nodata& \citet{Luck95}\\
4\,800 & 2.80 & 1.6 & \nodata&\nodata & +0.02$\pm$0.05&+0.12$\pm$0.05 & +0.31$\pm$0.05& \nodata& \nodata& \nodata& \citet{Ottmann98}\tablenotemark{b}\\
4\,850 & 2.40 & 1.6 & +0.48$\pm$0.2&\nodata & \nodata&+0.24$\pm$0.2 & \nodata& \nodata& \nodata& \nodata& {J.~Drake}\tablenotemark{c}\\
\hline\\
31 Com&&&&&&&&&&& \\\\
{T$_{eff}$}&
{log g}&
{$\xi$}&
{$[$Na/H$]$}&
{$[$Al/H$]$}&
{$[$Mg/H$]$}&
{$[$Fe/H$]$}&
{$[$Si/H$]$}&
{$[$S/H$]$}&
{$[$O/H$]$}&
{$[$Ne/H$]$}&
{References}\\\\
\hline
5\,550 & 3.00 & 1.7  & \nodata&\nodata & \nodata&-0.05$\pm$0.10 & \nodata& \nodata&\nodata & \nodata& \citet{Gustafsson74}\\
5\,680 & 3.00 & 1.7  & \nodata&\nodata & \nodata&-0.11$\pm$0.10 & \nodata& \nodata&\nodata & \nodata& \citet{Taylor99}\\
\hline\\
$\mu$ Vel&&&&&&&&&&& \\\\
{T$_{eff}$}&
{log g}&
{$\xi$}&
{$[$Na/H$]$}&
{$[$Al/H$]$}&
{$[$Mg/H$]$}&
{$[$Fe/H$]$}&
{$[$Si/H$]$}&
{$[$S/H$]$}&
{$[$O/H$]$}&
{$[$Ne/H$]$}&
{References}\\\\
\hline
5\,180 & 2.60 & \nodata &  \nodata&\nodata & \nodata&-0.07 & \nodata& \nodata&\nodata & \nodata& \citet{Burnashev83}\\
\hline
\end{tabular}
\end{center}
\tablenotetext{a}{Note that all the photospheric abundances are scaled to the values given by \citet{Asplund04}}
\tablenotetext{b}{We list here the formal "internal" accuracy stated by \citet{Ottmann98}. As these authors indicate, the true 
uncertainties will be significantly higher owing to systematic 
uncertainties.}
\tablenotetext{c}{J.~Drake unpublished, based on spectra obtained at McDonald Observatory 82'' telescope and echelle spectrograph.}
}
\end{table}

\newpage

\begin{deluxetable}{lrrr}
\tabletypesize{\scriptsize}
\tablecaption{Summary of {\it{Chandra}} HETG+ACIS-S Observations.\label{t:obs}}
\tablewidth{0pt}
\tablehead{\colhead{Star}  &\colhead{ObsID} & \colhead{Start [UT]} & \colhead{Exp [ks]}
} 
\startdata
31 Com &  1891 & 2001-03-12T15:03:38& 130.2 \\
$\mu$ Vel &  3410 & 2001-12-18T19:23:14 &  57.0 \\
$\mu$ Vel   &  1890 & 2001-09-24T08:35:16 & 19.7 \\
$\beta$ Cet	&  974 & 2001-06-29T23:29:11& 86.1 \\
\enddata
\end{deluxetable}

\newpage

\begin{sidewaystable}
\caption{Identification and fluxes for spectral lines, observed on $\beta$ Cet, 31 Com, $\mu$ Vel (ObsID 3410 and ObsID 1890), used in this analysis.\label{t:flx}}
\scriptsize{
\begin{center}
\begin{tabular}{lllcccccl}
\hline  \hline \\
{$\lambda_{\rm obs}$} & 
{$\lambda_{\rm pred}$} &
{Ion} & 
{$log\,T_{\rm max}$} & 
{31 Com} &
{$\mu$ Vel (ObsID 3410)} &
{$\mu$ Vel (ObsID 1890)} & 
{$\beta$ Cet}  & 
{Transition} \\
{[\AA]} & 
{[\AA]} &
{} & 
{[K]} & 
{[10$^{14}$\,erg\,cm$^{-2}$\,s$^{-1}$]}  & 
{[10$^{14}$\,erg\,cm$^{-2}$\,s$^{-1}$]} &
{[10$^{14}$\,erg\,cm$^{-2}$\,s$^{-1}$]} &
{[10$^{14}$\,erg\,cm$^{-2}$\,s$^{-1}$]}&(upper $\rightarrow$ lower) \\\\
\hline \\
     4.722 &      4.727 & S  XVI  & 7.40 & $0.32 \pm 0.35$ &$	0.45 \pm    0.42$&\nodata&$   1.85 \pm   1.09$  	       &{\small $(2p) \; ^2P_{3/2}$ $\rightarrow$ $(1s) \; ^2S_{1/2}$}  	     \\
\nodata &      4.733 & S     XVI  & 7.40 & $0.16 \pm 0.17$ &$	0.23 \pm    0.21$&\nodata &$   0.93 \pm  0.54$  	       &{\small $(2p) \; ^2P_{1/2}$ $\rightarrow$ $(1s) \; ^2S_{1/2}$}  	     \\
     5.032 &      5.039 & S  XV   & 7.20 & $1.83 \pm 0.69$ &$	1.35 \pm    1.22$&\nodata &$   8.37 \pm  1.51$  	       &{\small $(1s2p) \; ^1P_{1}$ $\rightarrow$ $(1s^2) \; ^1S_{0}$}  	     \\
     6.177 &      6.180 & Si XIV  & 7.20 & $3.77 \pm 0.33$ &$	3.15 \pm    0.52$&$   1.04 \pm    1.13$ &$   8.28 \pm	 0.59$ &{\small $(2p) \; ^2P_{3/2}$ $\rightarrow$ $(1s) \; ^2S_{1/2}$}  	     \\
\nodata &      6.186 & Si    XIV  & 7.20 & $1.88 \pm 0.16$ &$	1.57 \pm    0.26$&$   0.52 \pm    0.57$ &$   4.14 \pm	 0.29$ &{\small $(2p) \; ^2P_{1/2}$ $\rightarrow$ $(1s) \; ^2S_{1/2}$}  	     \\
     6.647 &      6.648 & Si XIII & 7.00 & $5.70 \pm 0.44$ &$	9.42 \pm    0.83$&$   9.04 \pm    1.56$ &$  20.74 \pm	 0.96$ &{\small $(1s2p) \; ^1P_{1}$ $\rightarrow$ $(1s^2) \; ^1S_{0}$}  	     \\
     7.176 &      7.176 & Al XIII & 7.10 & $0.14 \pm 0.04$ &$	0.21 \pm    0.21$&\nodata&$   0.18 \pm    0.09$ 	       &{\small $(2p) \; ^2P_{1/2}$ $\rightarrow$ $(1s) \; ^2S_{1/2}$}  	     \\
\nodata &      7.171 & Al    XIII & 7.10 & $0.27 \pm 0.09$ &$	0.43 \pm    0.43$&\nodata&$   0.36 \pm    0.18$ 	       &{\small $(2p) \; ^2P_{3/2}$ $\rightarrow$ $(1s) \; ^2S_{1/2}$}  	     \\
     7.732 &      7.757 & Al XII  & 6.90 & $0.82 \pm 0.30$ &$	1.64 \pm    0.51$&\nodata&$   2.06 \pm    0.49$ 	       &{\small $(1s2p) \; ^1P_{1}$ $\rightarrow$ $(1s^2) \; ^1S_{0}$}  	     \\
     8.422 &      8.425 & Mg XII  & 7.00 & $1.64 \pm 0.15$ &$	2.85 \pm    0.29$&$   2.37 \pm    0.57$ &$   7.85 \pm	 0.28$ &{\small $(2p) \; ^2P_{1/2}$ $\rightarrow$ $(1s) \; ^2S_{1/2}$}  	     \\
\nodata &      8.419 & Mg    XII  & 7.00 & $3.28 \pm 0.30$ &$	5.70 \pm    0.58$&$   4.75 \pm    1.14$ &$  15.72 \pm	 0.56$ &{\small $(2p) \; ^2P_{3/2}$ $\rightarrow$ $(1s) \; ^2S_{1/2}$}  	     \\
     9.173 &      9.169 & Mg XI   & 6.80 & $2.30 \pm 0.29$ &$  10.06 \pm    0.79$&$   9.37 \pm    1.43$ &$  22.30 \pm	 0.93$ &{\small $(1s2p) \; ^1P_{1}$ $\rightarrow$ $(1s^2) \; ^1S_{0}$}  	     \\
    10.028 &     10.029 & Na XI   & 6.90 & $0.29 \pm 0.10$ &$	1.26 \pm    0.30$&$   1.67 \pm    0.87$ &$   2.08 \pm	 0.24$ &{\small $(2p) \; ^2P_{1/2}$ $\rightarrow$ $(1s) \; ^2S_{1/2}$}  	     \\
\nodata &     10.023 & Na    XI   & 6.90 & $0.57 \pm 0.19$ &$	2.52 \pm    0.60$&$   3.33 \pm    1.74$ &$   4.16 \pm	 0.48$ &{\small $(2p) \; ^2P_{3/2}$ $\rightarrow$ $(1s) \; ^2S_{1/2}$}  	     \\
    10.998 &     11.003 & Na X    & 6.70 & $2.33 \pm 0.36$ &$	2.77 \pm    0.95$&$   1.49 \pm    1.38$ &$   7.94 \pm	 0.90$ &{\small $(1s2p) \; ^1P_{1}$ $\rightarrow$ $(1s^2) \; ^1S_{0}$}  	     \\
    12.133 &     12.132 & Ne X    & 6.80 & $6.24 \pm 0.46$ &$  13.37 \pm    0.88$&$  11.51 \pm    1.55$ &$  37.49 \pm	 1.12$ &{\small $(2p) \; ^2P_{3/2}$ $\rightarrow$ $(1s) \; ^2S_{1/2}$}  	     \\
\nodata &     12.137 & Ne    X    & 6.80 & $3.12 \pm 0.23$ &$	6.68 \pm    0.44$&$   5.75 \pm    0.77$ &$  18.73 \pm	 0.56$ &{\small $(2p) \; ^2P_{1/2}$ $\rightarrow$ $(1s) \; ^2S_{1/2}$}  	     \\
    12.283 &     12.285 & Fe XXI  & 7.00 & $7.70 \pm 0.64$ &$	9.71 \pm    1.03$&$   5.25 \pm    1.56$ &$  24.98 \pm	 1.25$ &{\small $(2s^22p^3d) \; ^3D_{1}$ $\rightarrow$ $(2s^22p^2) \; ^3P_{0}$}      \\
    13.433 &     13.447 & Ne IX   & 6.60 & $1.75 \pm 0.53$ &$	9.49 \pm    1.43$&$   9.56 \pm    2.94$ &$  20.07 \pm	 1.63$ &{\small $(1s2p) \; ^1P_{1}$ $\rightarrow$ $(1s^2) \; ^1S_{0}$}  	     \\
    13.508 &     13.504 & Fe XIX  & 6.90 & $3.76 \pm 0.41$ &$	9.06 \pm    0.99$&$   7.69 \pm    2.02$ &$  23.14 \pm	 1.16$ &{\small $(2p^3(^2P)3d) \; ^1D_{2}$ $\rightarrow$ $(2s^22p^4) \; ^3P_{2}$}    \\
    14.208 &     14.208 & Fe XVIII& 6.90 & $1.82 \pm 0.29$ &$	8.99 \pm    0.85$&$   6.83 \pm    1.48$ &$  20.66 \pm	 0.94$ &{\small $(2p^4(^1D)3d) \; ^2P_{3/2}$ $\rightarrow$ $(2s^22p^5) \; ^2P_{3/2}$}\\
\nodata &     14.203 & Fe    XVIII& 6.90 & $3.43 \pm 0.56$ &$  17.01 \pm    1.61$&$  12.91 \pm    2.80$ &$  39.07 \pm	 1.78$ &{\small $(2p^4(^1D)3d) \; ^2D_{5/2}$ $\rightarrow$ $(2s^22p^5) \; ^2P_{3/2}$}\\
    14.263 &     14.267 & Fe XVIII& 6.90 & $0.65 \pm 0.23$ &$	3.26 \pm    0.88$&$   3.73 \pm    2.44$ &$   7.60 \pm	 1.09$ &{\small $(2p^4(^1D)3d) \; ^2F_{5/2}$ $\rightarrow$ $(2s^22p^5) \; ^2P_{3/2}$}\\
    15.013 &     15.015 & Fe XVII & 6.75 & $8.32 \pm 1.12$ &$  63.06 \pm    4.21$&$  63.37 \pm    7.70$ &$ 129.30 \pm	 3.60$ &{\small $(2p^53d) \; ^1P_{1}$ $\rightarrow$ $(2p^6) \; ^1S_{0}$}	     \\
    16.008 &     16.007 & O  VIII & 6.50 & $1.59 \pm 0.24$ &$	3.89 \pm    0.50$&$   3.08 \pm    0.78$ &$   9.19 \pm	 0.56$ &{\small $(3p) \; ^2P_{1/2}$ $\rightarrow$ $(1s) \; ^2S_{1/2}$}  	     \\
\nodata &     16.006 & O     VIII & 6.50 & $3.17 \pm 0.47$ &$	7.75 \pm    1.00$&$   6.14 \pm    1.55$ &$  18.31 \pm	 1.12$ &{\small $(3p) \; ^2P_{3/2}$ $\rightarrow$ $(1s) \; ^2S_{1/2}$}  	     \\
    18.973 &     18.973 & O  VIII & 6.50 & $4.90 \pm 0.63$ &$  13.67 \pm    1.60$&$  12.72 \pm    3.13$ &$  30.38 \pm	 1.56$ &{\small $(2p) \; ^2P_{1/2}$ $\rightarrow$ $(1s) \; ^2S_{1/2}$}  	     \\
\nodata &     18.967 & O     VIII & 6.50 & $9.80 \pm 1.27$ &$  27.37 \pm    3.20$&$  25.45 \pm    6.26$ &$  60.80 \pm	 3.12$ &{\small $(2p) \; ^2P_{3/2}$ $\rightarrow$ $(1s) \; ^2S_{1/2}$}  	     \\
    21.607 &     21.602 & O  VII  & 6.30 & $3.12 \pm 1.83$ &$	6.04 \pm    4.60$&$   4.00 \pm    8.89$ &$   7.72 \pm	 2.69$ &{\small $(1s2p) \; ^1P_{1}$ $\rightarrow$ $(1s^2) \; ^1S_{0}$}  	     \\
    24.777 &     24.779 & N  VII  & 6.30 & \nodata	   &$	7.05 \pm    2.73$&\nodata&$  62.99 \pm   5.1 $  	        &{\small $(2p) \; ^2P_{3/2}$ $\rightarrow$ $(1s) \; ^2S_{1/2}$} 	     \\
\nodata &     24.785 & N     VII  & 6.30 & \nodata	   &$	3.52 \pm    1.36$&\nodata&$  31.48 \pm   2.6 $  	        &{\small $(2p) \; ^2P_{1/2}$ $\rightarrow$ $(1s) \; ^2S_{1/2}$} 	     \\
\hline
\end{tabular}
\end{center}
}
\end{sidewaystable}

\newpage

\begin{table}
\caption{Coronal abundances obtained from abundance-independent DEM-reconstructions.\label{t:abund1}}
\scriptsize{
\begin{center}
\begin{tabular}{lccccc}
\hline  \hline \\
Element\tablenotemark{a}& 
FIP\tablenotemark{b}&
31 Com&
$\mu$ Vel &
$\mu$ Vel &
$\beta$ Cet\\
& 
&
&
(ObsID 3410)&
(ObsID 1890)&\\
\hline \\
Na/H&5.14 & $<$+0.79\tablenotemark{c}& $<$+1.04\tablenotemark{c}& $<$+0.81\tablenotemark{c} &  $<$+0.93\tablenotemark{c}\\
Al/H&5.98 & $<$-0.02\tablenotemark{c}& $<$+0.20\tablenotemark{c}&    	     	    \nodata &  $<$+0.08\tablenotemark{c}\\
Mg/H&7.65 &    -0.12$\pm$	 0.04&    +0.08$\pm$	    0.03&    +0.11$\pm$ 	0.08&	  -0.01$\pm$	    0.02\\
Fe/H&7.87 &    -0.21$\pm$	 0.04&    -0.03$\pm$	    0.03&    -0.06$\pm$ 	0.07&	  -0.12$\pm$	    0.01\\
Si/H&8.15 &    -0.02$\pm$	 0.03&    +0.09$\pm$	    0.04&    +0.05$\pm$ 	0.11&	  -0.05$\pm$	    0.02\\
S/H &10.36&    -0.20$\pm$	 0.16&    -0.15$\pm$	    0.31&    	     	     \nodata&	  +0.06$\pm$	    0.08\\
O/H &13.62&    -0.11$\pm$	 0.08&    -0.18$\pm$	    0.07&    -0.27$\pm$ 	0.16&	  -0.20$\pm$	    0.03\\
Ne/H&21.56&    -0.37$\pm$	 0.04&    -0.41$\pm$	    0.04&    -0.45$\pm$        0.08 &	  -0.40$\pm$	    0.02\\\\
\hline
\end{tabular}
\end{center}
\small{$^a$Logarithmic abundances relative to the abundance mixture of \citet{Asplund04} with Ne from \citet{Drake05b} (see Table~\ref{t:photabun} and discussion in \S5.2).}\\
\small{$^b$First Ionization Potential in eV.}\\
\small{$^c$Upper limits due to lack of signal in line features.}\\
}
\end{table}

\begin{table}
\caption{$\beta$ Cet, 31 Com, $\mu$ Vel Abundance Ratios using Temperature-Insensitive Diagnostics.\label{t:abund2}}
\begin{center}
\begin{tabular}{lcccccccc}
\hline  \hline \\
Abundance&
31 Com&
31 Com\tablenotemark{a}&
$\mu$ Vel&
$\mu$ Vel\tablenotemark{a}&
$\mu$ Vel&
$\mu$ Vel\tablenotemark{a}&
$\beta$ Cet&
$\beta$ Cet\tablenotemark{a}\\
Ratio&
&&
(ObsID 3410)&
(ObsID 3410)&
(ObsID 1890)&
(ObsID 1890)&\\
\hline \\
$$N/O$$  &\nodata	    & \nodata        &+0.68$\pm$0.19&\nodata	   &  \nodata	  &\nodata	 &+1.32$\pm$0.03  &\nodata\\
$$O/Ne$$ & +0.22$\pm$	0.21& +0.26$\pm$0.12 &+0.07$\pm$0.13&+0.23$\pm$0.11&+0.05$\pm$0.18&+0.18$\pm$0.24&+0.06$\pm$ 0.13 &+0.20$\pm$0.05\\
$$Ne/Mg$$& -0.25$\pm$	0.06& -0.25$\pm$0.08 &-0.50$\pm$0.04&-0.49$\pm$0.07&-0.53$\pm$0.08&-0.56$\pm$0.16&-0.38$\pm$ 0.02 &-0.39$\pm$0.04\\
$$Ne/Fe$$& +0.17$\pm$	0.07& -0.16$\pm$0.08 &-0.34$\pm$0.04&-0.38$\pm$0.07&-0.40$\pm$0.08&-0.39$\pm$0.15&-0.22$\pm$ 0.02 &-0.28$\pm$0.03\\
$$Mg/Si$$& -0.15$\pm$	0.05& -0.10$\pm$0.07 &-0.04$\pm$0.05&-0.01$\pm$0.07&-0.07$\pm$0.11&+0.06$\pm$0.19&+0.03$\pm$ 0.02 &+0.04$\pm$0.04\\
$$Mg/Fe$$& -0.13$\pm$	0.07& +0.09$\pm$0.08 &+0.06$\pm$0.05&+0.11$\pm$0.06&+0.15$\pm$0.13&+0.17$\pm$0.15&-0.01$\pm$  0.03&+0.11$\pm$0.03\\
$$Si/S$$ & +0.18$\pm$	0.16& +0.18$\pm$0.19 &+0.27$\pm$0.34&+0.24$\pm$0.35&   \nodata    &\nodata	 &-0.10$\pm$ 0.08 &-0.11$\pm$0.10 \\\\

\hline
\end{tabular}
\end{center}
\small{$^a$Abundance ratios using from the DEM-reconstructions.}\\
\end{table}

\newpage

\newpage

\newpage


\newpage

\end{document}